\documentclass[a4paper]{mem}
\usepackage{natbib}
\usepackage{graphicx}
\usepackage[a4paper]{hyperref}

\idline{XX}{1}

\def\about{$\sim$}
\def\Figure#1#2[#3] {\centering \leavevmode \includegraphics[width=#3,clip]{#1.ps}}

\def\r                {\hbox{$r$}}
\def\i                {\hbox{$i$}}

\def\ug               {\hbox{$u-g$}}
\def\gr               {\hbox{$g-r$}}
\def\ri               {\hbox{$r-i$}}

\begin{document}

     \title{          Variability Studies with SDSS             }

\author{\v{Z}. Ivezi\'{c} \inst{1},
 R.H. Lupton \inst{1},
 S. Anderson \inst{2}
 L. Eyer \inst{1},
 J.E. Gunn \inst{1}, \\
 M. Juri\'c \inst{1},
 G.R. Knapp \inst{1},
 G. Miknaitis \inst{2}, 
 J.E. Gunn \inst{1}, 
 C.M. Rockosi \inst{2}, \\
 D. Schlegel \inst{1}, 
 M.A. Strauss \inst{1},
 C. Stubbs \inst{2}
 \and 
 D.E. Vanden Berk \inst{2}
}

   \mail{ivezic@astro.princeton.edu}

   \institute{$^1$Princeton University, $^2$University of Washington, $^3$University of Pittsburgh
             }

   \abstract{
   The potential of the Sloan Digital Sky Survey for wide-field variability
   studies is illustrated using multi-epoch observations for 3,000,000 point sources 
   observed in 700 deg$^2$ of sky, with time spans ranging from 3 hours to 3 years. 
   These repeated observations of the same sources demonstrate that SDSS delivers 
   $\sim$0.02 mag photometry with well behaved and understood errors.
   We show that quasars dominate optically faint ($r \ga 18$) point sources that are 
   variable on time scales longer than a few months, while for shorter time scales, 
   and at bright magnitudes, most variable sources are stars.
   }
  
   \authorrunning{\v{Z}. Ivezi\'{c} et al.}
   \titlerunning{Variability Studies with SDSS}
   \maketitle
%

\section{Introduction}

The properties of optically faint variable sources are by and large
unknown. There are about 10$^9$ stars brighter than $V=20$ in the sky,
and at least 3\% of them are expected to be variable at a few percent
level (Eyer, 1999). However, the overwhelming majority are not recognized
as variables even at the brightest magnitudes: 90\% of the variable
stars with $V<12$ remain to be discovered (Paczy\'{n}ski, 2000).

The Sloan Digital Sky Survey will significantly contribute to studies
of optically faint variable sources due to its accurate multi-epoch 
photometry for a large sky area (currently 1,000 deg$^2$, and up to
4,000 deg$^2$ by the survey completion). Here we present a preliminary
variability analysis for 3,000,000 point sources detected in 700 deg$^2$ 
of sky, and multiply observed with time scales ranging from 3 hours to 
3 years.

\begin{figure*}[]
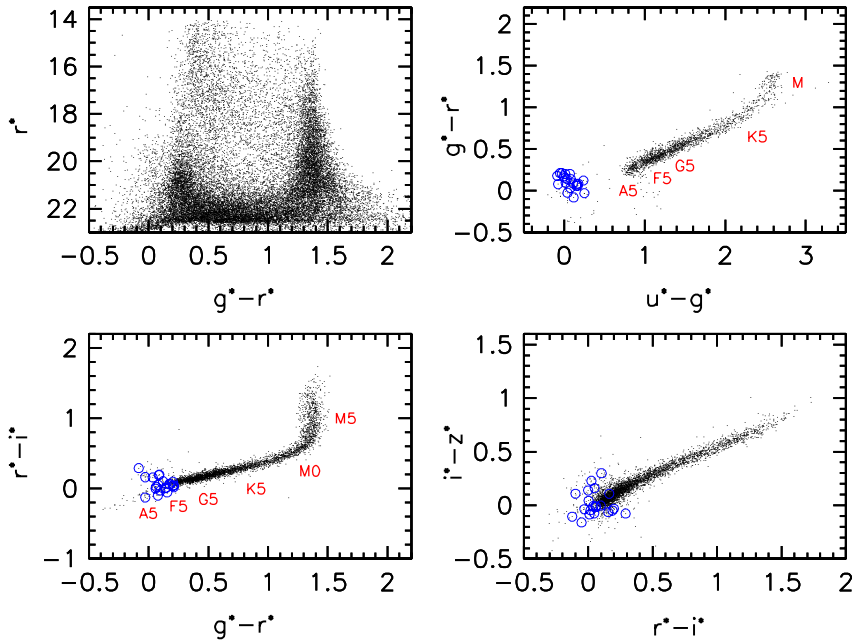

\centering 
\begin{minipage}{\textwidth}
\vskip -1.6in
\hskip -0.8in
\Figure{SDSSpsfCMD}{ps}[1.1\hsize]
\vskip -2.7in
\caption{The SDSS color-color and color-magnitude diagrams which summarize
photometric properties of unresolved sources, marked as dots. The top left panel
displays the \r\ vs. \gr\ color-magnitude diagram for \about 25,000 objects observed
in 3 deg$^2$ of sky. The three remaining panels show
color-color diagrams for objects brighter than 20$^m$ in each of the 3 bands used
to construct each diagram (red is always towards the upper right corner).
The locus of ``normal" stars is clearly visible in all three diagrams,
and the positions of several spectral types are indicated next to the locus in
the \ug\ vs. \gr\ and \ri\ vs. \gr\ color-color diagrams (labels are slightly offset
for clarity). Objects that have \ug\ and \gr\ colors similar to those of low-redshift 
($z < 2$) QSOs (\ug $<$ 0.4, -0.1 $<$ \gr $<$ 0.3,
\ri $<$ 0.5), and are brighter than the limit for quasars in the SDSS spectroscopic
survey (\i $<$ 19), are marked by open circles.}
\label{SDSSall}
\end{minipage}
\end{figure*}

\begin{figure*}[]
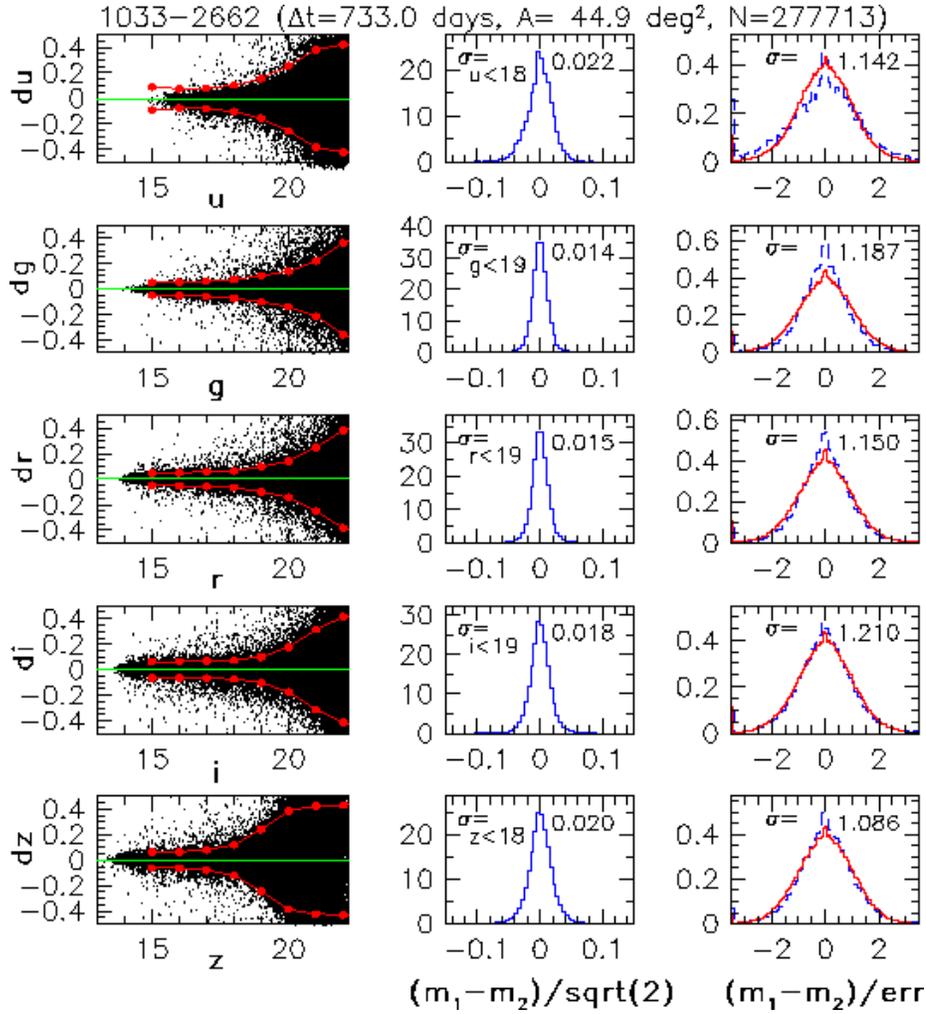

\centering 
\begin{minipage}{\textwidth}
\vskip 0.5in
\Figure{photoQA}{ps}[0.9\hsize]
\bigskip
\caption{Analysis of the SDSS photometric accuracy using repeated observations
(733 days apart) of 280,000 unresolved sources detected in 45 deg$^2$
of sky. The five rows correspond to the five SDSS bands. The panels in the first
column show the difference between the two measurements as a function of
apparent (psf) magnitude. The big dots, connected by the lines to guide the eye,
are the 3$\sigma$ envelope determined from the interquartile range in 1 mag
wide bins. The middle column shows the histograms of magnitude difference
divided by $\sqrt{2}$ (a quantity representative of mean photometric errors)
for the bright end, with the limiting magnitude shown in each panel.
The equivalent Gaussian width of these distributions, determined from the
interquartile range, is also shown in each panel. The last column
displays the distribution of magnitude differences normalized by the expected
formal errors. The two barely distinguishable histograms correspond to
bright sources, and to the entire sample. Their mean equivalent Gaussian
width, $\sigma$, is indicated in each panel. Note that the values of $\sigma$
are \about 1, testifying that the formal errors computed by the photometric 
pipeline are highly accurate.}
\label{photoQA}
\end{minipage}
\end{figure*}

\begin{figure}[t]
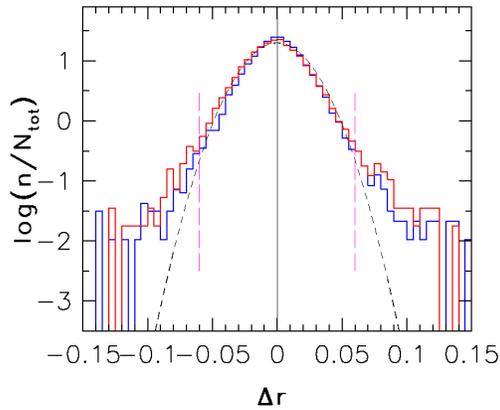

\Figure{tails2}{ps}[1.0\hsize]
\caption{The thick solid histograms show the distribution of the measured 
$r$ band difference for two color-selected subsamples of stars observed 
twice 3 hours apart (note the logarithmic scale!). A Gaussian with the same 
interquartile range, corresponding to $\sigma$=0.02 mag, is shown by the 
short-dashed line. As evident, the distribution of SDSS photometric errors 
is almost a perfect Gaussian.
\label{tails}
}
\end{figure}

\subsection{  The Basic Characteristics of the SDSS Imaging Survey}

The Sloan Digital Sky Survey (SDSS; York et al.~2000; Stoughton et al. 
2002) is providing homogeneous and deep ($r < 22.5$) photometry in five 
passbands ($u$, $g$, $r$, $i$, and $z$, Fukugita et al.~1996; Gunn et 
al.~1998) accurate to 0.02 mag, of up to 10,000 deg$^2$ in the Northern 
Galactic Cap, and a smaller, but deeper, survey of 200 deg$^2$ in the 
Southern Galactic Hemisphere. The survey sky coverage will result in 
photometric measurements for about 100 million stars and a similar number 
of galaxies. Astrometric positions are accurate to better than 0.1 arcsec 
per coordinate (rms) for sources brighter than 20.5$^m$ (Pier et al.~ 2003), 
and the morphological information from the images allows robust star-galaxy 
separation to $\sim$ 21.5$^m$ (Lupton et al.~2001).

\subsection{   The SDSS Multi-epoch Observations }

SDSS imaging data are obtained by tracking the sky in six parallel
scanlines, each 13.5 arcmin wide. The six scanlines from two nights
are then interleaved to make a filled stripe. Because of the scan
overlaps, and because of the scan convergence near the survey poles,
about 40\% of the sky in the northern survey will be surveyed
twice. In addition, all of the southern survey areas will be observed
dozens of times to search for variable objects and, by stacking the
frames, to go deeper. 

While two observations are normally insufficient to characterize a
variable object, the multi-color nature and accuracy of the SDSS
photometric data helps enormously.
For example, the $u$ band even allows remarkably efficient selection
of the low-metallicity G and K giants (Helmi et al. 2002) and
blue horizontal branch stars (Yanny et al. 2000).

\subsection{The Colors of Point Sources in the SDSS Photometric System }

The position of an unresolved source in multi-dimensional SDSS color space
is an excellent proxy for its classification. For example, the spectral type 
for most stars can be estimated to within 2 spectral subtypes, or better.
Furthermore, for quasars, which can be efficiently selected by their non-stellar 
colors (Richards et al. 2002), it is possible to obtain a reasonably accurate 
photometric redshift (Budavari et al. 2001, Richards et al. 2001).

The SDSS color-color and color-magnitude diagrams which summarize photometric
properties of unresolved sources are shown in Figure \ref{SDSSall}. 
The position of a star in these diagrams is mainly determined by its spectral 
type. The modeling of the stellar populations observed by SDSS indicates 
that the vast majority of these stars (about 99\%) are on the main sequence
(Finlator {\em et al.} 2000).

\subsection{  How Accurate are the SDSS Data?   }

\begin{figure*}[]
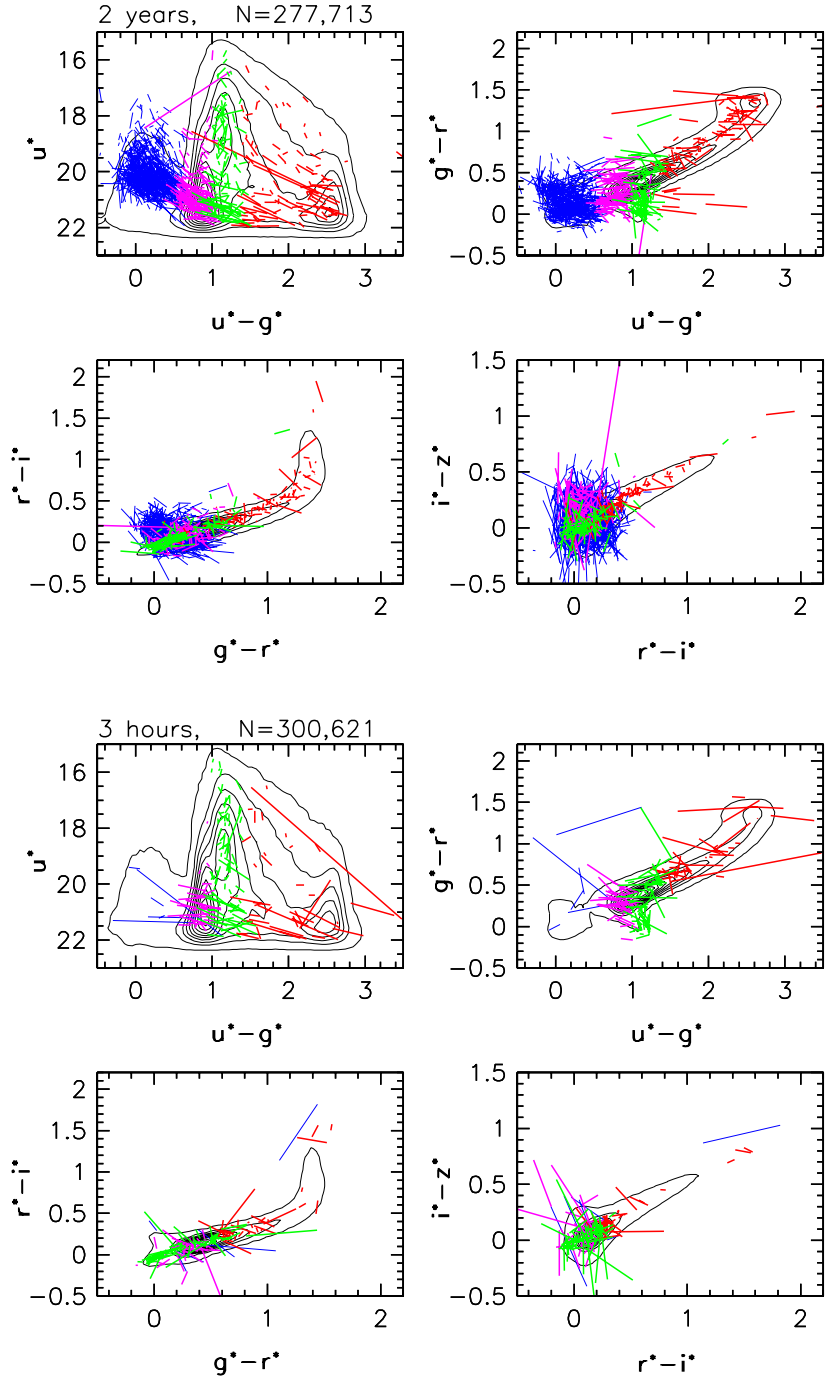

\centering 
\begin{minipage}{\textwidth}
\vskip -0.2in
\Figure{plotClasses_2years}{ps}[0.85\hsize]
\vskip -2.3in
\Figure{plotClasses_3hours}{ps}[0.85\hsize]
\vskip -2.3in
\caption{The top four panels show variable sources discovered 
in 75 deg$^2$ of sky observed twice 733 days apart. The overall source 
distribution is shown as contours, using the mean magnitudes. The two 
measurements for variable objects are connected by lines. The bottom
four panels show variable sources discovered in observations obtained
3 hours apart. Note the absence of variable quasars ($u-g<0.6$) in the
bottom four panels.}
\label{longT}
\end{minipage}
\end{figure*}

SDSS offers unprecedented photometric accuracy for such a large scale optical
survey. Not only are the photometric errors generally small, but they
themselves are accurately determined by the photometric pipeline ({\em photo},
Lupton et al. 2001), and can be reliably used to estimate the statistical
significance of measured magnitude differences. This ability is of paramount importance
for a robust statistical study of variable objects. We demonstrate these claims
by comparing the measurements for \about 280,000 unresolved sources from 45 deg$^2$
of sky observed twice 733 days apart.

The five rows of panels in Figure \ref{photoQA} correspond
to five SDSS bands. The first column shows the difference between the two
measurements as a function of apparent magnitude. At the bright end the errors
are roughly independent of magnitude (because they are dominated by an imperfect
description of the point-spread function), and then increase towards the
faint end. As the panels in the middle column show, the internal accuracy of SDSS 
photometry is about 0.02 mag or better in all five bands.
The SDSS is currently the only survey that provides multi-epoch imaging
data to a limit as faint as \r \about 22, in an area as large as 1000 deg$^2$, 
in five bands spanning the wavelengths from UV to IR, with such small errors.

A distinctive feature of SDSS photometry is the well controlled
tails of the photometric error distribution. Figure \ref{tails} shows
the magnitude difference distribution for stars brighter than $r=19$
from 32 deg$^2$ of sky observed twice 3 hours apart. The measured 
distribution closely follows a Gaussian with $\sigma$=0.02 mag, 
all the way to $\pm 3\sigma$. There are only 0.9\% observations outside 
the $\pm 3\sigma$ range, in good agreeement with the value expected for a 
perfect Gaussian (0.3\%). Of course, some of these objects are certainly
variable and thus increase the fraction of $\pm 3\sigma$ outliers.

\subsection{  What Twinkles in the Faint Optical Sky ?    }

Using multiple observations of 3,000,000 point sources detected 
in 700 deg$^2$ of sky, we selected variable objects by requiring a minimum 
variation of at least 0.075 mag in both $g$ and $r$ bands, and statistical 
significance of at least $3 \sigma$. The variable population strongly
depends on the time difference between the two observations, and
also on the object's magnitude: for sufficiently long time scales
(a few months or longer), variable objects fainter than r \about 18 are dominated 
by quasars. 
For time scales shorter than about a month, or at bright magnitudes, 
the variable objects are heavily dominated by stars.
    
Figure \ref{longT} shows the color-magnitude and color-color diagrams analogous 
to those shown in Figure \ref{SDSSall}, except that here the overall source 
distribution is shown as contours, using the mean magnitudes, while the two 
measurements for variable objects are connected by lines. The top four panels
correspond to observations obtained 733 days apart, and the bottom four
panels to observations obtained 3 hours apart. A good handle on the types of 
detected variable sources can be obtained by simply studying their $u-g$ color 
distribution. For a $\sim$2 year time scale, the majority of objects (77\%) have 
colors typical of low-redshift quasars ($u-g < 0.6$). Blue stars consistent 
with the halo turn off, both in colors and in apparent magnitude, make up 40\% 
of the remaining objects (this subsample may also contain some quasars), 
and the reddish stars with $u-g>1.3$ another 40\%. RR Lyrae stars, which can 
be easily recognized thanks to their distinctive colors (Ivezi\'{c} et al. 2000), 
comprise 20\% of detected variable stars. At the 3 hours time scale, 
only about 2\% of variable objects have quasar colors. RR Lyrae now make 
up 35\% of the sample, mainly because the number of red stars is much smaller 
(the surface density of selected RR Lyrae is about 1 per deg$^2$, nearly 
independent of time scale and galactic coordinates). The smaller number of 
red stars is consistent with them being long-period (Mira) variables which 
don't appreciably vary on time scales of several hours.

For a more detailed description of this work please see
Ivezi\'{c} et al. and Vanden Berk et al. (in prep.).

\begin{acknowledgements}
Funding for the creation and distribution of the SDSS Archive has been provided 
by the Alfred P. Sloan Foundation, the Participating Institutions, the National 
Aeronautics and Space Administration, the National Science Foundation, the U.S.
Department of Energy, the Japanese Monbukagakusho, and the Max Planck Society. 
The SDSS Web site is http://www.sdss.org/. 

The SDSS is managed by the Astrophysical Research Consortium (ARC) for the 
Participating Institutions. The Participating Institutions are The University of 
Chicago, Fermilab, the Institute for Advanced Study, the Japan Participation Group, 
The Johns Hopkins University, Los Alamos National Laboratory, the Max-Planck-Institute 
for Astronomy (MPIA), the Max-Planck-Institute for Astrophysics (MPA), New Mexico 
State University, University of Pittsburgh, Princeton University, the United States 
Naval Observatory, and the University of Washington.
\end{acknowledgements}

\bibliographystyle{aa}

\end{document}